\documentclass[12pt,notoc]{JHEP3}

\usepackage[latin1]{inputenc}
\usepackage[T1]{fontenc}
\usepackage[dvips]{graphicx}
\usepackage{amsmath} \usepackage{amsfonts} \usepackage{amssymb}
\usepackage{bbm}
\DeclareMathOperator{\tgh}{tgh}
\DeclareMathOperator{\tr}{Tr\,}
\DeclareMathOperator{\supp}{supp}

\title{The Coulomb branch of the Leigh-Strassler deformation and matrix models}

\author{Francesco Benini\\ International School for Advanced Studies (SISSA)\\ Via Beirut 2-4, I-34013 Trieste, Italy\\
and Scuola Normale Superiore\\ Piazza dei Cavalieri 7, Pisa I-56126, Italy\\ E-mail: \email{f.benini@sns.it}}

\abstract{
The Dijkgraaf-Vafa approach is used in order to study the Coulomb branch of the Leigh-Strassler massive deformation of $\mathcal{N}=4$ SYM with gauge group $U(N)$.
The theory has $\mathcal{N}=1$ SUSY and an $N$-dimensional Coulomb branch of vacua, which can be described by a family of ``generalized'' Seiberg-Witten curves.
The matrix model analysis is performed by adding a tree level potential that selects particular vacua. The family of curves is found: it consists of order $N$ branched coverings of a base torus, and it is described by multi-valued functions on the latter. The relation between the potential and the vacuum is made explicit. The gauge group $SU(N)$ is also considered.
Finally the resolvents from which expectation values of chiral operators can be extracted are presented.
}

\preprint{hep-th/0411057}


\begin{document}

\newcommand{\valmed}[1]{\langle #1 \rangle}
\newcommand{\Valmed}[1]{\left\langle #1 \right\rangle}
\newcommand{\partfrac}[2]{\frac{\partial #1}{\partial #2}}
\newcommand{\sub}[2]{#1_\text{#2}}
\newcommand{\re}{\,\Re e \:}
\newcommand{\im}{\,\Im m \:}
\newcommand{\virg}[1]{``#1''}
\newcommand{\hyph}[1]{\mbox{$#1$}\nobreakdash-\hspace{0pt}}
\newcommand{\thetauno}[2]{\, \theta_1 \Bigl[ #1 \Big| #2 \Bigr]}
\newcommand{\map}[3]{#1 \negmedspace : \negmedspace #2 \negmedspace \mapsto \negmedspace #3}
\newcommand{\setmeno}{\negmedspace \smallsetminus \negmedspace}

\newcommand{\Nugual}[1]{$\mathcal{N}= #1 $}
\newcommand{\xmeno}{$\{ x=-\infty \}$}
\newcommand{\xpiu}{$\{ x=+\infty \}$}
\newcommand{\xpiumeno}{$\{ x=\pm\infty \}$}
\newcommand{\zzero}{$\{ z=0 \}$}

\def\Xint#1{\mathchoice
    {\XXint\displaystyle\textstyle{#1}}%
    {\XXint\textstyle\scriptstyle{#1}}%
    {\XXint\scriptstyle\scriptscriptstyle{#1}}%
    {\XXint\scriptscriptstyle\scriptscriptstyle{#1}}%
    \!\int}
\def\XXint#1#2#3{{\setbox0=\hbox{$#1{#2#3}{\int}$}
	\vcenter{\hbox{$#2#3$}}\kern-.5\wd0}}
\def\dashint{\Xint\diagup}


\section{Introduction}

The Dijkgraaf-Vafa (DV) conjecture \cite{dijkgraaf} has proven a powerful 
tool for studying  
the low energy limit of  \Nugual{1} gauge theories with
massive chiral fields.
The method consists in solving an auxiliary matrix model, 
where the \Nugual{1} chiral fields are replaced by random matrices 
interacting with a potential equal to the tree level superpotential. 
This method allows to extract the holomorphic quantities of 
the low energy regime, 
including all the non-perturbative effects. 
The procedure has been tested for a lot of different theories \cite{dijkgraaf,anomalies,phases}. When solving the matrix model, an interesting geometrical structure emerges, which is characterized by a Riemann surface provided with a meromorphic differential. This structure is based on special geometry and
generalizes the Seiberg-Witten curve for \Nugual{2} theories. 
As in the \Nugual{2} case, all the holomorphic quantities in the effective
Lagrangian can be derived from the geometry. For \Nugual{1} theories, this
amounts to the determination of the effective superpotential and the 
kinetic terms for low energy massless photons.
The expectation values of the chiral operators can be also determined
from some meromorphic differentials defined 
on the surface \cite{anomalies,phases}.

The DV conjecture can be also used to derive the Seiberg-Witten 
curve for \Nugual{2} theories. If we deform the  \Nugual{2} theory with a 
superpotential for the adjoint field, we select particular points in moduli
space. With a sufficiently generic potential, we can explore the entire
moduli space. As the tree level potential is adiabatically removed, 
the matrix model Riemann surface runs in the Seiberg-Witten curve 
of the \Nugual{2} theory. This argument has been introduced in \cite{fluxes} 
and later used for  
theories with different matter content and in different dimensions
 \cite{zaffaroni,vafaholl}.
The very same argument can be used to study \Nugual{1} theories 
with a Coulomb branch of vacua. The matrix model description determines a
generalized Seiberg-Witten curve, whose periods determine the coupling constant
matrix. Clearly, the curve does not determine the entire effective Lagrangian
as in the \Nugual{2} case, but only the holomorphic quantities. 
A particularly interesting model in this respect is the so-called 
Leigh and Strassler (LS) deformation of \Nugual{4} SYM \cite{leigh}. It is 
known that there exist superconformal deformations of \Nugual{4} SYM 
which preserve \Nugual{1} SUSY. In particular, 
the LS theory has the same fields 
as \Nugual{4} SYM and a superpotential
\begin{equation} \label{eq28}
\mathcal{W}_\textrm{tree} = \tr \Big( i\lambda \Phi[\Phi^+,\Phi^-]_\beta  \Big) \:,
\end{equation}
where the \hyph{\beta}commutator is defined by
$$
[\Phi^+,\Phi^-]_\beta \equiv \Phi^+ \Phi^- e^{i\beta/2} - \Phi^- \Phi^+ e^{-i\beta/2} \:.
$$
Here $\lambda$ is a function of $\beta$, which is tuned in order to preserve 
superconformal invariance. 
Fortunately we will not need its explicit expression. We will always
consider an \Nugual{2}* type theory where two adjoints fields are made
massive by the addition of the superpotential $M\Phi^+\Phi^-$.

The theory has a great varieties of classical vacua \cite{sdualityof,deconstruction,exact,criticalpoints}. We will mainly be interested in the Coulomb phase: $\Phi^\pm=0$. The Coulomb branch of the theory  will be described by a 
family of generalized SW curves, which encodes the dynamics. The form of 
the curve for the LS model was conjectured in \cite{hollowood}, using 
integrable system arguments. It is a deformation of the Donagi-Witten
curve for \Nugual{2}* \cite{dw}. Here we derive the curve using the DV prescription. Our strategy is to 
analyze the model in presence of a  deformation $\sub{W}{tree}(\Phi)= \sum_{k=1}^{n+1} \frac{g_k}{k} \Phi^k$. It is known \cite{fluxes,transition} that 
such deformation lifts the moduli space, constraining the vacua on particular \hyph{(N-n)}dimensional sub-varieties where $(N-n)$ monopoles are massless. 
The eigenvalues of $\Phi$ distribute themselves over the $n$ critical points of $W$, with multiplicity $N_i$. The gauge group is then broken to $U(N) \to \prod_{i=1}^n U(N_i)$. 
At the quantum level, other non-perturbative effects join. The $SU(N_i)$ factors confine, 
leaving the group $U(1)^n$.%
\footnote{Unless some $N_i$ vanish, in which case more monopoles condense and less photons are left.}
At low energy, we can write an effective potential in terms of the chiral superfields $S_i = \tr_{U(N_i)} (W_\alpha W^\alpha)$, and this 
can be computed through the matrix model. 
As we will see, its value is just the expectation 
value of the tree level superpotential
$$
\sub{W}{eff} = \valmed{\sub{W}{tree}(\Phi)} \:,
$$
after the latter has selected the vacuum. We can study the Coulomb phase of the LS model  by taking a 
degree $N+1$ potential $\sub{W}{tree}$, and considering the vacua characterized by $N_i=1$. 
In this way, we can explore the whole moduli space by varying the potential.

\

The matrix model has already been solved \cite{sdualityof,kostov}. The vacuum with gauge group $\prod_{i=1}^n U(N_i)$ is associated with an \hyph{n}cut solution, which generates a Riemann surface of genus $n$. The structure is very similar to that in \cite{zaffaroni}: the eigenvalue space in the large matrix limit is a cylinder with pairs of cuts identified. The physical curve is obtained  
upon minimization and  it becomes the covering of a torus. 
We find the explicit form of the curve: $F(v,z)=0$,
 a degree $n$ polynomial in $v$, whose coefficients 
are multi-valued function of the base torus coordinate $z$ with determined monodromy properties. 
This expression was firstly conjectured in \cite{hollowood}. 
We also identify the moduli for gauge groups $U(N)$ and $SU(N)$.%
\footnote{Observe that the LS model does not show a complete decoupling of the gauge multiplet 
associated with $\tr \Phi$. Actually, as it can be seen from the Lagrangian, 
the vector multiplet $(A_\mu,\lambda_\alpha)$ decouples, 
while the chiral one $(\tr \Phi, \psi_\alpha)$ does not.}
Finally, we identify the resolvents which 
allow to evaluate expectation values of operators in the chiral ring.


\section{The matrix model and the Riemann surface}

The $U(N)$ gauge theory with \Nugual{1} SUSY can be studied through the planar limit solution of the associated matrix model, as suggested by Dijkgraaf e Vafa \cite{dijkgraaf}. The matrix model is defined by the partition function
\begin{align*}
Z &= e^{-\mathcal{F}/g_s^2} \\
&= \int d\hat\Phi d\hat\Phi^+ d\hat\Phi^- \exp \left\{ -\frac{1}{g_s} \tr \Big( i\lambda \hat\Phi[\hat\Phi^+,\hat\Phi^-]_\beta + M \hat\Phi^+ \hat\Phi^- + W_\textrm{tree}(\hat\Phi) \Big) \right\} \:,
\end{align*}
where $\mathcal{F}$ is the free energy and $\hat \Phi_i$ are $\hat N\times \hat N$ matrices. 
The right implementation of the group $U(N)$ is realized taking $\hat \Phi$ hermitian whereas $\hat \Phi^+ = (\hat \Phi^-)^\dag$ \cite{dijkgraaf}.

We consider the 't Hooft limit in which the dimension $\hat N$ of the matrices goes to infinity and the coupling $g_s$ goes to zero, while keeping fixed the product $S = g_s \hat N$. The associated matrix model has already been solved in another contest \cite{kostov}. The procedure is the following: first integrate out $\hat \Phi^\pm$, then scale $\hat \Phi \to \hat \Phi/\lambda$ and change variables to get an integral over the eigenvalues $\phi_i$ of $\hat \Phi$.
Eventually make another change of variables to $\delta_i$ and define a new function $V(\delta)$ :
\begin{equation} \label{eq02}
\phi_i \equiv \frac{M}{\lambda} \left( -e^{\delta_i} + \frac{1}{2\sin \beta/2} \right) \qquad \qquad V(\delta) \equiv W\Big( -\frac{M}{\lambda}e^\delta + \frac{M}{2\lambda \sin \beta/2} \Big) \:. \end{equation}

In the saddle point approximation for large $\hat N$ one can get the equations of motion
\begin{equation} \label{eq01}
V'(\delta_k) = \frac{g_s}{2} \sum_{j(\neq k)} \left\{ \frac{2}{\tgh \frac{\delta_k-\delta_j}{2}} - \frac{1}{\tgh \frac{\delta_k-\delta_j+i\beta}{2}} - \frac{1}{\tgh \frac{\delta_k-\delta_j-i\beta}{2}} \right\}
\end{equation}
As usual, one can distribute the eigenvalues over the $n$ critical points of the potential $V(x)$ with multiplicities $\hat N_i$. The continuum limit is studied through the eigenvalues density function $\rho(x)$ and the resolvent function
\begin{equation} \label{eq30}
\omega(x) = \frac{1}{2\hat N} \sum_j \frac{1}{\tgh \dfrac{x-\delta_j}{2}} \qquad \to \qquad \frac{1}{2} \int \frac{\rho(y)}{\tgh \dfrac{x-y}{2}} dy \:.
\end{equation}
In this limit the filling fractions $S_i=g_s \hat N_i$ parametrize the particular solution around which we are expanding, and they are identified with the condensate chiral fields. According to the DV prescription we can obtain the effective superpotential from the expression
$$
\sub{W}{eff} = \sum_{i=1}^n N_i \partfrac{\mathcal{F}}{S_i} + 2\pi i \tau S
$$
where $N_i$ indicate the particular vacuum $\prod_{i=1}^n U(N_i)$ we are considering, while $\mathcal{F}(S_i)$ is the matrix model free energy in the planar limit.

It is useful to introduce the function
\begin{equation} \label{eq29}
G(x) \equiv U(x) + i S \Big[ \omega(x+i\frac{\beta}{2}) - \omega(x-i\frac{\beta}{2}) \Big] \:,
\end{equation}
where $U(x)$ is a degree $n+1$ polynomial in $e^x$ defined by
\begin{equation} \label{eq13}
V'(x) = -i \Big[ U(x+i\frac{\beta}{2}) - U(x-i\frac{\beta}{2}) \Big] \:.
\end{equation}
The saddle point equation \eqref{eq01} takes the simple form
\begin{equation} \label{eq03}
G(x+i\frac{\beta}{2}\pm i\epsilon) = G(x-i\frac{\beta}{2}\mp i\epsilon) \qquad x\in \supp \rho \:.
\end{equation}

\

The holomorphic change of variables \eqref{eq02} maps the complex plane into the cylinder (the strip $\{-i\pi \leq \im x \leq i\pi\}$ with the two sides identified). We can add the two points at infinity \xpiu{} and \xmeno{} in order to compactify it.
$U(x)$ is a degree $n+1$ polynomial in $e^x$, so that it is a well-defined meromorphic function on the cylinder, regular at \xmeno{} and with a pole of order $n+1$ at \xpiu; $\omega(x)$ has $n$ cut discontinuities on the support of $\rho(x)$ while it is regular at extremes. Therefore $G(x)$ presents $n$ pairs of cuts $[a_i^- \pm i\beta/2, a_i^+ \pm i\beta/2]$, and a pole of order $n+1$ at \xpiu. Eq. 
\eqref{eq03} tell us that $G(x)$ is a well-defined meromorphic function on a Riemann surface obtained identifying the two cuts of every pair.

\

Let us introduce a canonical basis of compact cycles on the Riemann surface: $A_i$ cycles circling
 the upper cuts and $B_i$ cycles going from the lower cuts to the upper ones. 
By integrating eq. \eqref{eq29} along \hyph{A_i}periods we extract the quantities
\begin{equation} \label{eq04}
S_i = \frac{1}{2\pi} \oint_{A_i} G(x)dx \:.
\end{equation}
The quantities $\partial\mathcal{F}/\partial S_i$ can be extracted from \hyph{B_i}periods \cite{dijkgraaf}, with a minor subtlety \cite{sdualityof}. By rescaling $\hat\Phi \to \hat\Phi/\lambda$, 
we gain a factor $\lambda^{-\hat N^2}$ that brings a further $S$ dependence. We have
$$
\mathcal{F} = S^2 \log \lambda + \tilde{\mathcal{F}}
$$
where $\tilde{\mathcal{F}}$ is the contribution from our saddle point integral. Therefore we get
$$
\partfrac{\mathcal{F}}{S_i} = 2S\log \lambda -i \oint_{B_i} G(x) dx \:.
$$

Finally we can put the superpotential in the form
\begin{equation} \label{eq05}
\sub{W}{eff} = -i \sum_{i=1}^n \left[ N_i \oint_{B_i} G \:dx - \Big( \tau-i\frac{N}{\pi} \log \lambda \Big) \oint_{A_i} G \: dx \right] \:.
\end{equation}
Note that $\sub{W}{eff}$ only depends on the \emph{renormalized coupling constant} \cite{sdualityof}
$$
\tau_R \equiv \tau - i\frac{N}{\pi} \log \lambda \:.
$$

\

Let us briefly analyze the geometrical structure, as in \cite{zaffaroni}. The function $G(x)$ has a single pole of order $n+1$ at \xpiu, and is meromorphic on the Riemann surface $\Sigma$. By Riemann-Roch, 
this requirement fixes the function up to a multiplicative constant and an additive one. The multiplicative constant is also fixed by comparison with $U(x)$ (i.e. with the potential $V(x)$), and the additive one is physically irrelevant.
$G(x)$ is therefore uniquely determined.

The differential $dx$ is well-defined on the Riemann surface: it has two simple poles at \xpiumeno{}, and it has periods
$$
\oint_{A_i} dx = 0 \qquad \qquad \oint_{B_i} dx = i\beta \qquad \qquad \oint_D dx = 2\pi i \:,
$$
Here the $D$ cycle wraps the cylinder and encircle, say, the \xpiu{} puncture. 
Let us count the number of moduli. We have $3n-3$ parameters from the moduli space of 
Riemann surfaces of genus $n$, plus $2$ from the punctures. By Riemann-Roch, there  exist $n+1$ 
meromorphic differentials with two poles at the punctures; the integration to get the $x$ coordinate 
involves another additive constant. Taking into account  that $2n+1$ periods are fixed we obtain
$$
(3n-1) + (n+2) - (2n+1) = 2n \:.
$$

From the matrix model point of view, we can interpret these as the classical moduli space of the $n$ roots of $V'(x)$ plus the $n$ fields $S_i$. Or, alternatively, 
 as the $2n$ zeros of the differential $dx$, prescribed by Riemann-Roch, 
which coincides with the extremes of the cuts $[ a_i^- \pm i\beta/2, a_i^+ \pm i\beta/2 ]$.

\section{The on-shell theory}

We have written the effective potential as a function of the classical vacuum around which we are expanding (characterized by $N_i$), and a function of the condensate fields $S_i$. The true quantum vacua are obtained by minimizing with respect to $S_i$.

\

By varying the potential with respect to the field $S_i$ \cite{zaffaroni} we get the differentials
\begin{equation} \label{eq07}
\omega_i = \frac{1}{2\pi} \partfrac{}{S_i} G(x)dx \:.
\end{equation}
They constitute a canonical basis of independent holomorphic differentials on $\Sigma$, associated to cycles. In particular from \eqref{eq04} we have $\oint_{A_i} \omega_j = \delta_{ij}$.

From the minimizing of \eqref{eq05} we get
\begin{equation} \label{eq08}
\oint_{B_i} \sum_j N_j \omega_j \equiv \oint_{B_i} \Omega = \tau_R \qquad \qquad \forall i \:,
\end{equation}
where we used the symmetry of the period matrix $\oint_{B_j} \omega_i$. The equation tell us that, on-shell, it exists on the surface a meromorphic form $\Omega \equiv \sum_j N_j \omega_j$ with periods 
\begin{equation} \label{eq12}
\oint_{A_i} \Omega = N_i \qquad \qquad \oint_{B_i} \Omega = \tau_R \:.
\end{equation}

We can integrate $\Omega$ to get a well-defined function $z(P) = \int_{P_0}^{P} \Omega$ 
on a torus defined by the identifications
$$
z \equiv z + \tilde N \equiv z + \tau_R \:,
$$
where $\tilde N$ is the greatest common factor of the $N_i$. So we have demonstrated that, on-shell, $\Sigma$ is the covering of a torus of modular parameter $\tau_R/\tilde N$\phantom{ }%
\footnote{Note that this is equivalent to the covering of a torus of modular parameter $\tau_R$, but with an order $\tilde N$ times greater.}.
Furthermore, the images of the periods $A_i$ and $B_i$ wind around the torus $N_i/\tilde N$ and $1$ times respectively.

\subsection{The curve}

Our aim is now to obtain an expression for the curve $\Sigma$ of the on-shell theory. We know that it is the covering of a torus, so, in analogy to what happens for \Nugual{2^*} theories \cite{dw,phong2}, we can write it as
$$
F(v,z) = \prod_{i=1}^{\tilde n} [v-v_i(z)] \:,
$$
where $v_i(z)$ describe the sheets, while $\tilde n$ is the order (till unknown) of the covering.

In analogy with \cite{zaffaroni}, we would like to construct a meromorphic function $v$ on $\Sigma$, with simple poles at the $\tilde n$ counter-images $p_i$ of the point \zzero, and such that the form $v\,dz$ has residues $1$ at $\tilde n-1$ of them. Since the function $G(x)$ creates a punctures, it is
convenient to fix the base point $P_0$ at the point at infinity \xpiu, so that $P_0$ 
is one of the counter-images of the point \zzero.

\

To construct the function $v(x)$ it is better to start by its differential. It should have second order 
poles at the counterimages of  \zzero.
The idea is to consider a meromorphic differential on the base torus with a double pole in \zzero{} and coefficient $-1$, so that its pull-back on $\Sigma$ has $\tilde n$ double poles with equal coefficients, 
and then correct the pole at $P_0$ with a differential on $\Sigma$ written in terms of $x$. We can choose
$e^x dx$. Unlike $dx$, it is multi-valued%
\footnote{It is worth to observe that, from Riemann-Roch, a meromorphic form on the surface with exactly a double pole exists unique. We don't know its explicit expression, but in general it will not be integrable.}; in going from a lower cut to the upper one it takes a phase: $x \to x+i\beta$, $e^x dx \to e^{i\beta} e^x dx$.
Therefore we construct $v$ with the same property:
\begin{equation} \label{eq06}
\begin{array}{rclcl}
v & \to & v & \qquad & \text{along $A_i$ cycles} \\ 
v & \to & e^{i\beta}v & \qquad & \text{along $B_i$ cycles} \:.
\end{array} 
\end{equation}

Now we can construct multi-valued functions on the base torus through the function
\begin{equation} \label{eq15}
k(z) \; = \; \frac{\sigma \big( z-\frac{\omega_1\beta}{\pi} \big)} {\sigma(z) \sigma \big( -\frac{\omega_1\beta}{\pi} \big)} \; e^{\beta \eta_1 z/\pi} \; = \; \frac{\pi}{2\omega_1} \frac{\theta_1'[0] \frac{ }{ }} {\theta_1 \bigl[ -\frac{\beta}{2} \bigr]} \frac{\theta_1 \bigl[ \frac{\pi z}{2\omega_1} - \frac{\beta}{2} \bigr]} {\theta_1 \bigl[ \frac{\pi z}{2 \omega_1} \bigr]} \:,
\end{equation}
where $2\omega_1 = \tilde N$ and $2\omega_2=\tau_R$ are the torus periods, and we have used semi-elliptic functions. Actually, $k(z)$ has a simple pole at \zzero{} with residue $1$, a simple zero at $\{z=\omega_1 \beta / \pi\}$, and satisfies the expected monodromy properties.

\

Finally, we can write down the differential of the function $v$
$$
dv = \eta\, e^x dx + k(z) \partfrac{}{z} \log k(z) \, dz
$$
and, by integrating, we get
\begin{equation} \label{eq16}
v = \eta\, e^x + k(z) \:.
\end{equation}
Note that the required monodromy excludes any integration constant, while $\eta$ can be reabsorbed by a translation in $x$.

\subsection{Order of the covering}
\label{coveringorder}

The surface $\Sigma$ is the covering of a base torus, realized through the function $z$. Now we determine the order of the covering, that is the number of sheets.
To this purpose we use  the following residue theorem for meromorphic forms with the prescribed 
monodromy \eqref{eq06}:%
\footnote{This is proved easily by ''flattening'' the surface on the plane: every compact orientable Riemann surface of genus $g$ is 
homeomorphic to a complex polygon  with $4g$ sides, and with suitable identifications on the boundary. In particular, we can identify sides according to the string
$ \alpha_1 \cdot \beta_1 \cdot \alpha_1^{-1} \cdot \beta_1^{-1} \cdot \dotsm \cdot \alpha_g \cdot \beta_g \cdot \alpha_g^{-1} \cdot \beta_g^{-1} $. For $\alpha_i$ and $\beta_i$ we can take the cycles $A_i$ and $B_i$. Then it suffices to deform the path that encircles the poles into the perimeter of the polygon, and sum the contributions.
We also notice that every function with the monodromy \eqref{eq06} is expressible through $k(z)$ and a generic elliptic function:
$ f(z) = g(z) \cdot k(z). $
Moreover, the only function with a single pole in \zzero{} is exactly $k(z)$.}

\begin{equation} \label{eq09}
\sum_p \: \text{Res}_p \: \omega = \frac{e^{-i\beta}-1}{2\pi i} \sum_i \oint_{A_i} \omega \:.
\end{equation}
Consider now the expression
$$
\oint_{A_i} v\,dz = \oint_{A_i} \eta \, e^x\,dz + \frac{N_i}{\tilde N} \oint_A k(z) dz \:.
$$
The differential $dz$%
\footnote{With $dz$ we will indifferently refer to the differential on the torus and its pull-back on $\Sigma$.}
is defined by \eqref{eq08} and \eqref{eq07}:
$$
dz = \sum_{j=1}^N \frac{1}{2\pi} N_j \partfrac{}{S_j} G(x)\,dx \:.
$$
We can convert the integral of $G(x)dx$ along the cycle $A_i$ into the integral of the eigenvalues density $\rho(x)$ over the segment of support $[a_i^-,a_i^+]$, by using the relation
$$
\big[ -G(x+i\epsilon) + G(x-i\epsilon) \big] = 2\pi S \: \rho \bigl( x-i\frac{\beta}{2} \bigr) \qquad \qquad x \in A_i \:.
$$
We now use the residue theorem \eqref{eq09} for $v\,dz$ and for the function $k(z)$ on the torus separately, and sum over the cycles. Defining the quantities $W_n \equiv \int_{\mathbb{R}} \rho(x) e^{nx} dx$, we get
\begin{equation}\label{eqnew}
\sum_p \: \text{Res}_p \:(v\,dz) = - \eta\, \frac{e^{i\beta/2}-e^{-i\beta/2}}{2\pi i} \sum_{j=1}^N \partfrac{}{S_j} [S\,W_1] + \frac{N}{\tilde N} \:.
\end{equation}

\

On the other hand, we can evaluate the residues by encircling the poles of $v$ ($dz$ is an holomorphic form). The poles are located over the point \zzero, and are as many as the sheets of the covering. Consider the expression
$$
\sum_{i=1}^{\tilde n} \: \text{Res}_{p_i} \: (v\,dz) = \text{Res}_{+\infty} \: \eta \, e^x\,dz + \sum_{i=1}^{\tilde n} \: \text{Res}_{\,0} \: k(z)dz \:.
$$
$k(z)$ has unitary residues and we deal with $e^x dz$ by expanding around the point at infinity:
$$
G(x)dx \quad \to \quad U(x)dx + 2S\sum_{m=1}^\infty \sin \frac{m\beta}{2} W_m e^{-mx} dx \:.
$$
Eventually we obtain
\begin{equation} \label{eq21}
\sum_p \: \text{Res}_p \:(v\,dz) = - \frac{\eta}{\pi} \sin \frac{\beta}{2} \sum_{j=1}^N \partfrac{}{S_j} [S\,W_1] + \tilde n \:.
\end{equation}

By comparison with expression \eqref{eqnew}, 
we deduce the order of the covering: \mbox{$\tilde n = N/\tilde N$}, in 
agreement with what we have been expecting from the analogy with the \Nugual{2^*} case \cite{zaffaroni}.

\subsection{Curve and factorization}
\label{curve}

We have found the multi-valued coordinate system $(v,z)$ over $\Sigma$, where $z$ establishes an \hyph{\tilde n}covering of the base torus. Therefore the equation of the curve is an order $\tilde n$ polynomial in $v$ whose coefficients are multi-valued functions on the torus:
$$
0 = F(v,z) = \sum_{m=0}^{\tilde n} F_m(z) \, v^{\tilde n-m} \qquad \qquad F_0 \equiv 1 \:.
$$
The functions $F_m(z)$ are such that going along the \hyph{B}cycle of the torus they take a phase $e^{im\beta}$. So the function $F(v,z)$ is multi-valued too, and takes a phase $e^{i\tilde n\beta}$ going 
along the \hyph{B}cycle. This result agrees with \cite{hollowood}, where it was obtained through integrable systems.

We can also write the local expression
\begin{equation} \label{eq14}
0=F(v,z) = \prod_{i=1}^{\tilde n} \big[ v-v_i(z) \big] \:,
\end{equation}
where the ``functions'' $v_i(z)$ represent the counter-images of torus points: they are local coordinates over the covering sheets. Along the \hyph{B_i}cycles of $\Sigma$ they take a phase $e^{i\beta}$, and
they get permuted in going from a sheet to another one.

\

\begin{figure}[tbp]
\begin{center}
\includegraphics[width=\textwidth]{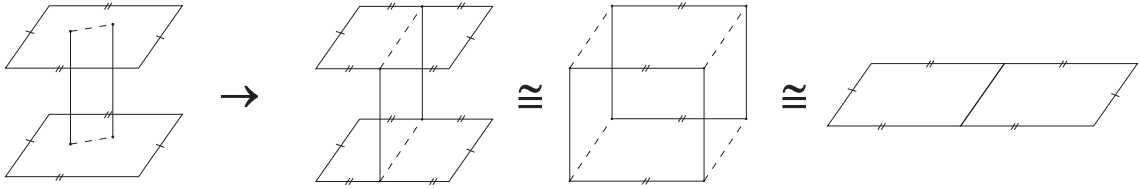}
\caption{\emph{Degeneration of the surface}. When two branching points collide, we can take the cut wound around a cycle, obtaining a torus of modified modular parameter.}
\label{cut_degeneration}
\end{center}
\end{figure}
In order to explore the whole moduli space, we need \cite{fluxes} to turn on a degree $N+1$ potential and to 
consider vacua characterized by $N_i=1$. In this case we find an \hyph{N}covering of a torus with modular parameter $\tau_R$ which gives the LS curve. 

We can also consider the case with generic $N_i$. For $N_i>1$, non-abelian gauge groups classically
unbroken undergo confinement. The superpotential has dynamically selected points in the moduli space of 
vacua of the LS theory where some monopoles are massless \cite{transition,monopole}. These are generically points where
the LS curve degenerates. The number of effective pairs of cuts on the cylinder (those with $N_i \geqslant 1$), 
that is the number of handles or the genus, and the number of covering sheets are related by
$$
\# \text{ pairs of cuts } = g_\Sigma = \sum_{i \text{ t.c. } N_i \geqslant 1} 1 \leqslant \sum_i \frac{N_i}{\tilde N} = \tilde n
$$
where equality holds just for $N_i=\tilde N$. The covering is generically singular, 
and $\Sigma$ is its desingularization. The number of non-coincident branching points 
$\nu$ of the function $v(z)$ is given by Riemann-Hurwitz formula:
$$
1+\nu/2 = g_\Sigma \leqslant \tilde n \:.
$$
Therefore, when a potential is applied, some branching points must collide, and their branching 
cuts will shrink. 
We can always take, even in the general case, a base torus of modular parameter $\tau_R$ and an \hyph{N}covering with some coincident branching points (see figure \ref{cut_degeneration}).

\

We end with the case of maximal degeneration, that is one $N_i=N$ while all the others null. 
In this vacuum the gauge group is not classically broken, but confinement only leaves a $U(1)$ factor. 
We have a \hyph{1}covering, that is a homeomorphism, of a torus of modular parameter $\tau_R/N$. The curve is $e^x = \gamma \, k(z)$. This perfectly agrees with \cite{sdualityof}.

\section{Resolvents}

We are now interested in operators in the chiral ring. 
The matrix model has been solved through the change of variables \eqref{eq02}:
\begin{equation*} 
\Phi = -\frac{M}{\lambda} e^\Delta + \frac{M}{2\lambda \sin \beta/2} \qquad \qquad W(\Phi) = V(\Delta)
\end{equation*}
from the field $\Phi$ to $\Delta$. This leads to the variable $x$, associated with the 
eigenvalues of $\Delta$, and which lives on the cylinder. It is useful to define the QFT resolvents
\begin{align*}
\mathcal{R}(x) &= -\frac{1}{32\pi^2} \biggl\langle \tr \frac{W_{\alpha i} W^\alpha_i}{2 \tgh \bigl( \frac{x-\Delta}{2} \bigr)} \biggr\rangle \\
\mathcal{T}(x) &= \biggl\langle \tr \frac{1}{2 \tgh \bigl( \frac{x-\Delta}{2} \bigr)} \biggr\rangle \:.
\end{align*}
The perturbative expansion around the point \xpiu{}
\begin{align*}
\mathcal{R}(x) &= -\frac{1}{32\pi^2} \Bigl\{ \frac{N}{2} + \sum_{k=1}^\infty e^{-kx} \langle \tr W_{\alpha i} W^\alpha_i e^{k\Delta} \rangle \Bigr\}\\
\mathcal{T}(x) &= \frac{N}{2} + \sum_{k=1}^\infty e^{-kx} \langle \tr e^{k\Delta} \rangle
\end{align*}
generates expectation values for the QFT chiral operators. 
It is clear that by knowing the expectation values of $e^{k\Delta}$ we can get all the expectation values of $\Phi^k$, because they are related by a linear transformation.

We expect that the resolvents can be  analytically continued and extended to the whole Riemann surface
as in \cite{anomalies,phases}. 
In the classical limit, the differentials $\mathcal{R}(x)dx$ and $\mathcal{T}(x)dx$ 
become meromorphic forms with simple poles at the eigenvalues of $\Delta$ and at the points \xpiumeno%
\footnote{Actually, in the classical limit the resolvent $\mathcal{R}$ vanishes, for it contains a fermionic bilinear; on the other hand, classically the fields $S_i$ are null.}.
In the quantum theory the singularities spread out in cuts. We can evaluate the periods around the cuts 
with the formula \cite{anomalies},
\begin{equation} \label{eq10}
\frac{1}{2\pi i} \oint_{C_i} dx\,M \frac{1}{2\tgh \Bigl(\dfrac{x-\Delta}{2} \Bigr)} = \tr M P_i \:,
\end{equation}
where $M$ is an operator, $C_i$ a cycle around the cut and $P_i$ the projector on the gauge subgroup $U(N_i)$. We obtain the periods
\begin{align*} 
\frac{1}{2\pi i} \oint_{A_i} \mathcal{R}(x) dx &= -\frac{1}{32\pi^2} \bigl\langle \tr W_\alpha W^\alpha P_i \bigr\rangle = S_i \nonumber \\
\frac{1}{2\pi i} \oint_{A_i} \mathcal{T}(x) dx &= \tr P_i = N_i \:.
\end{align*}
We will identify the field theory resolvents with suitable meromorphic forms in the matrix model by
comparing periods. A real proof of the following statements would require a detailed analysis of the Ward
identities in field theory, using the Konishi anomaly as done in \cite{anomalies}, which is hard for \Nugual{2}* type models.
Nevertheless, we believe that the following result holds.
The $\mathcal{R}$ resolvent is easily identified as \mbox{$ \mathcal{R}(x) \equiv S\omega(x) $}.
Indeed we have
$$
\langle \tr W_\alpha W^\alpha P_i \rangle = \frac{1}{2\pi i} \oint_{C_i} \mathcal{R}(x)dx = \frac{1}{2\pi i} \oint_{C_i} S \omega(x) dx = S_i \:,
$$
where we used \eqref{eq10} and $C_i$ is a cycle around the \hyph{i}th segment $[a_i^-,a_i^+]$ of the support of $\rho(x)$.

\

In order to identify the $\mathcal{T}$ resolvent, it is better to consider a quantity 
that admits an analytic extension on the Riemann surface $\Sigma$. We define \cite{zaffaroni} the form
$$
t(x)dx = \Bigl[ \mathcal{T}(x-i\tfrac{\beta}{2}) - \mathcal{T}(x+i\tfrac{\beta}{2}) \Bigr] dx \:.
$$
Analyzing its expansion at the points \xpiumeno{}, we see that it is regular (the poles cancel). 
The only singularities left are the cuts. We conjecture that, on-shell, $t(x)dx$ admits an analytic continuation on the whole surface, being an holomorphic form on $\Sigma$. As shown in \cite{phases,fluxes}, it is a general feature that on-shell an holomorphic form exists.
From \eqref{eq10} we easily get that the \hyph{A_i}periods of $t(x)dx$ are $N_i$. Since the form is holomorphic, its \hyph{A_i}periods specify it completely. 
We can identify it with the matrix model differential $dz$:
$$
dz \equiv \frac{1}{2\pi i} t(x)dx = \Bigl[ \mathcal{T}(x-i\tfrac{\beta}{2}) - \mathcal{T}(x+i\tfrac{\beta}{2}) \Bigr] dx \:.
$$
Observe that, on-shell, its \hyph{B_i}periods are $\tau_R$.

\subsection{Expectation values}

We can compute all of field theory expectation values $\valmed{e^{k\Delta}}$ by exploiting the general fact that cut discontinuities of the holomorphic differential (here $t(x)dx$) represent the quantum eigenvalues distribution.

Let us evaluate the integral of $t(x)dx$ multiplied by a power of $e^x$, along a path $G$ that encloses all the pairs of cuts. We name $A_i$ the cycles around the upper cuts, while $A_i^*$ the ones 
around the lower cuts; the considered path encircles both. We can proceed in two ways. 
By deforming the path into two circles around the points \xpiumeno{}, from the expansion of 
$\mathcal{T}(x)$, we get
$$
\frac{1}{2\pi i} \oint_G e^{kx} \, t(x)dx = 2i \sin \frac{k\beta}{2} \langle \tr e^{k\Delta} \rangle \:.
$$
On the other hand, we take advantage of the identification $t(x)dx \equiv 2\pi i\,dz$ and deform the path into the sum of the cycles $A_i$ and $A_i^*$, obtaining
$$
\oint_G e^{kx} \,dz = \sum_{i=1}^n \oint_{A_i+A_i^*} e^{kx} \,dz = 2i \sin \frac{k\beta}{2} \sum_{i=1}^n \oint_{A_i} e^{k(x-i\beta/2)} \,dz \:.
$$

By comparison:
\begin{equation} \label{eq11}
\langle \tr e^{k\Delta} \rangle = \sum_{i=1}^n \oint_{A_i} e^{k(x-i\beta/2)} \,dz \:.
\end{equation}
This coincides perfectly with the result in \cite{sdualityof}, in case of massive $SU(N)$ vacua with a single cut.

\

Note the difference between the eigenvalues distribution in the matrix model and that in field theory. The former is described by $\rho(x)$; the latter by $dz$.
Therefore we have
\begin{equation} \label{eq22}
\begin{array}{rcll}
W_k &=& \langle\langle \tr e^{k \widehat\Delta} \rangle\rangle_{MM} \qquad & \text{in the matrix model} \\
\displaystyle \sum_{j=1}^n N_j \partfrac{}{S_j} \bigl[ SW_k \bigr] &=& \langle \tr e^{k\Delta} \rangle_{QFT} & \text{in field theory} \:.
\end{array}
\end{equation}

\subsection{The effective potential}

At this point it is interesting to use \eqref{eq11} to evaluate the on-shell effective potential. Through \eqref{eq12} this is expressible in terms of resolvent periods:
$$
\sub{W}{eff} = -i \sum_{i=1}^n \Bigl\{ \oint_{A_i} dz \oint_{B_i} G\,dx - \oint_{B_i} dz \oint_{A_i} G\,dx \Bigr\} \:.
$$

\begin{figure}[tbp]
\begin{center}
\includegraphics[height=3.5cm]{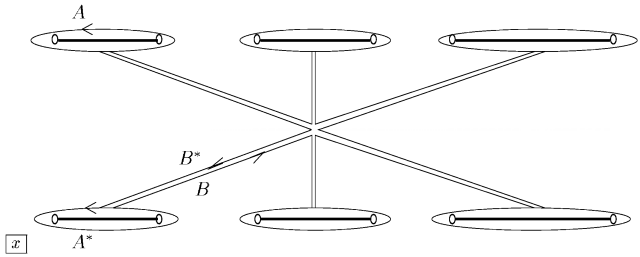}
\caption{\emph{Simply connected surface}. If we take into account the points at infinity, the system of cycles cuts off the surface and makes it simply connected.}
\label{sempl_connessa_zaffaroni}
\end{center}
\end{figure}
Now we perform a manipulation very similar to that in \cite{zaffaroni}. 
We cut off the surface, following figure \ref{sempl_connessa_zaffaroni}, in order to make it simply connected.%
\footnote{This procedure is equivalent to flatten the surface into a polygon with $4g$ sides. Note that we need not to eliminate the cycle $D$ around the cylinder, because $\oint_D dz=0$ and so $dz$ is already integrable.}
Then it suffices to deform the path from the cuts to the extremes of the cylinder, and back. Actually we have:
\begin{align*}
\sub{W}{eff} &= i \sum_{i=1}^n \oint_{A_i+A_i^*+B_i+B_i^*=G} z\,G\,dx \\
	&= i\oint_{-\infty} z\,G\,dx + i\oint_{+\infty} z\,G\,dx = i \oint_G z\,U\,dx \\
	&= \sum_{i=1}^n \left\{ \oint_{A_i} z \Bigl[ U(x)-U(x-i\beta) \Bigr] dx - \oint_{A_i} dz \oint_{B_i} U(x) dx \right\} \\
	&= \sum_{i=1}^n \oint_{A_i} V \bigl( x-i\tfrac{\beta}{2} \bigr) \, \frac{dz}{dx} \,dx -i N \int_{-i\beta/2}^{i\beta/2} U(x)\,dx -N\,V(0) \:.
\end{align*}
In the second line we observed that the only contribution at the extremes comes from $U(x)$. In the 
last line we used \eqref{eq13} and integrated by parts, 
making the $B_i$ cycles passing by two conventional points, say $\{ \pm i\beta/2 \}$.

Then we may replace $V$ with the original tree level potential $\sub{W}{tree}$, and use \eqref{eq11} to turn to expectation values. We get
$$
\sub{W}{eff} \sim \valmed{\tr \sub{W}{tree}(\Phi)} \:,
$$
up to additive constants which may depend on the potential, but not on the coupling $\tau$ or on the particular vacuum. So the effective superpotential 
is equal to the expectation value of the tree level potential.

\

The presence of additive constants is expected. We know \cite{exact,mixings} that a mixing ambiguity affects condensates in \Nugual{1^*} theories, and so also in the Leigh-Strassler model. The problem is that the operators $\Phi^n$ are ill-defined, and different approaches can lead to vacuum-independent mixings of them (including the identity).

\section{The curve}

The ``generalized'' Seiberg-Witten curve for the Leigh-Strassler model has been identified with
the matrix model curve, provided we take an order $N+1$ tree level potential and consider the vacuum 
characterized by $N_i=1$ \cite{fluxes}. In section \ref{curve} we showed that it may be written, as in \eqref{eq14}, through multi-valued functions on the base torus; this embeds $\Sigma$ into a twisted $\mathbb{C}$ bundle over $\mathbf{T}^2$. This expression hides the $N$ moduli which parameterize the moduli space. We want to bring the curve to a more suitable form, miming \cite{phong2}.

\

The idea is to subtract to $v$ a function with right monodromy properties, and which has a simple pole in \zzero{} with unit residue. We define a new polynomial function
\begin{align*}
f(v,z) &\equiv F(v+k(z),z) \\
&= \prod_{i=1}^N \bigl( v + k(z) - v_i(z) \bigr) = \sum_{m=0}^N f_m(z) \, v^{N-m} \qquad \text{with } f_0\equiv 1 \:,
\end{align*}
where $k(z)$ was introduced in \eqref{eq15}. The functions $f_m(v,z)$ have the same monodromies than 
$F_m(v,z)$, but the former have just a simple pole. This fixes them completely: they are proportional to
$$
\frac{ \theta_1 \bigl[ \frac{\pi z}{2\omega_1} - \frac{m \beta}{2} \big| \tilde\tau \bigr]}{ \theta_1 \bigl[ \frac{\pi z}{2\omega_1} \big| \tilde\tau \bigr] } \:,
$$
where $\tilde \tau$ is the modular parameter of the torus. Here $\tilde\tau=\tau_R$.

Reminding \eqref{eq16} for $v$, we can write the curve $0=F(v,z)$ in terms of the matrix model coordinate $x$:
\begin{equation} \label{eq17}
0 = \sum_{m=0}^N \gamma_m \, \thetauno{ \dfrac{\pi z}{2\omega_1} - m\dfrac{\beta}{2} }{\tilde\tau} \, e^{(N-m)x} \qquad \text{with } \gamma_0\equiv 1 \:.
\end{equation}
The parameters $\gamma_m$ are really the coordinates of the moduli space. It is convenient 
to normalize to $\gamma_0=1$, so that we are left with exactly $N$ parameters which describe 
the vacua for a $U(N)$ gauge group.

\

It is useful to find another expression for the curve, in order to analyze the semiclassical limit $g\to 0$ and compare with literature. We want to collect the $N$ parameters $\gamma_m$ in a single normalized degree $N$ polynomial in $e^x$:
$$
H(x) = \sum_{m=0}^N \gamma_m \, e^{(N-m)x} = \prod_{j=1}^N \bigl( e^x - e^{\xi_j} \bigr) \qquad \text{with } \gamma_0\equiv 1 \:.
$$
We apply the known expansion of the $\theta_1$ function:
$$
\theta_1[z|\tilde\tau] = -i \sum_{k \in \mathbb{Z}} \,(-1)^k \,q^{(k+\frac{1}{2})^2} \,e^{i(2k+1)z} \:,
$$
where $q=e^{i\pi\tilde \tau}$ contains the dependence on the modular parameter. With simple algebraic steps we get
\begin{equation} \label{eq18}
0 = \sum_{k\in\mathbb{Z}} \, (-1)^k \, q^{(k+\frac{1}{2})^2} \, e^{i(2k+1) {\textstyle [ \frac{\pi z}{2\omega_1} - \frac{N\beta}{2} ]} } \: H \bigl( x+i(2k+1)\tfrac{\beta}{2} \bigr) \:.
\end{equation}



\subsection{Gauge group $SU(N)$}

The curve for gauge group $SU(N)$ can be derived by enforcing the constraint $\valmed{\tr \Phi}=0$.
We use the curve written as in \eqref{eq14}: $F=\prod [v-v_i(z)]$. 
The multi-valued functions $v_i(z)$ have simple poles in \zzero{}, with $N-1$ residues equal to $1$ and
the last fixed by the residue theorem \eqref{eq09}. 
Using \eqref{eq21} and \eqref{eq22}, we get the following values for the residues
\begin{equation*} 
\Bigl( -\frac{\eta}{\pi} \sin \frac{\beta}{2} \langle \tr e^\Delta \rangle +1,1,\dotsc,1 \Bigr) \:.
\end{equation*}
In the polynomial function $f(v,z)$ the sum of the residues
appears in the term $v^{N-1}$. By comparison we obtain a relation between the parameter $\gamma_1$ and the field expectation value:
\begin{equation} \label{eq24}
\gamma_1 = - \frac{1}{2\omega_1} \frac{ \theta_1'[0] \sin \beta/2}{\theta_1[\beta/2]} \Bigl( -\frac{\lambda}{M} \langle \tr \Phi \rangle + \frac{N}{2 \sin \beta/2} \Bigr) \:.
\end{equation}

The family of SW curves that describes the moduli space for gauge group $SU(N)$ is obtained by freezing 
the parameter $\gamma_1$. So we are left with $N-1$ free parameters, whose number matches the rank of $SU(N)$.

\

The constraint $\valmed{\tr \Phi}=0$ may be applied to $F(v,z)$ too. We get the exact expression for the term $v^{N-1}$:
$$
F(v,z) \ni \left[ N - \frac{\eta N}{2\pi} + \frac{\eta\lambda}{\pi M} \sin \frac{\beta}{2} \valmed{\tr \Phi} \right] k(z) v^{N-1} \:.
$$
With the particular choice $\eta=2\pi/2\omega_1$%
\footnote{Which corresponds to a shift in the variable $x$.},
we cancel the term. This agrees with what stated in \cite{hollowood}, on the basis of integrable systems. Note that this holds also for generic vacua, in which case we must accomplish the substitutions
$$
N \to \tilde n = \frac{N}{\tilde N} \qquad -\frac{\eta N}{2\pi} = -\frac{N}{2\omega_1} \to -\frac{N}{\tilde N} \:.
$$

\subsection{The function $G(x)$}

The function $G(x)$ is completely determined by $\sub{W}{tree}$. Geometrically, the reason is 
the following.
The saddle point equation determines the existence of the function $G(x)$, meromorphic on a genus $N$ 
Riemann surface obtained identifying pairs of cuts. $G(x)$ must have a pole of order $N+1$ on $\Sigma$, and Riemann-Roch tells us that such a function is unique, up to a multiplicative constant and an additive one. So the curve $\Sigma$ fixes $G(x)$, and vice versa. Moreover  the pole structure of $G(x)dx$ is 
determined by the polynomial $U(x)$, that is by the potential $\sub{W}{tree}$. Therefore, 
$\sub{W}{tree}$ fixes all the parameters $\gamma_m$.

\

We want to construct explicitly the function $G(x)$ on the surface. We must request that it has an \hyph{(N+1)}order pole at the puncture \mbox{$\{ x=+\infty, z=0 \}$} and that it is meromorphic.
To this purpose, we consider the elliptic functions
$$
e^{mx} \; \frac{ \theta_1 {\textstyle [ \frac{\pi z}{2\omega_1} + \frac{m\beta}{2} | \tilde\tau ]} }{ \theta_1 {\textstyle [ \frac{\pi z}{2\omega_1} | \tilde\tau ]} } \:,
$$
for $m=1,\dots,N$. They present an \hyph{(m+1)}order pole at $\{ x=+\infty, z=0 \}$ and $N-1$ simple poles at $\{x \text{ finite},z=0 \}$. Moreover they are single valued.

The idea is to construct $G(x)$ by assembling these $N$ functions, in such a way to cancel the $N-1$ simple poles at finite points, leaving just the pole at the puncture. It is a matter of solving a linear system, and counting variables and equations we see that the solution is determined except for a multiplicative constant (and an additive one for $G(x)$).

Let us analyze the behavior of the curve in a neighborhood of \zzero. Expanding \eqref{eq17} we get
\begin{equation} \label{eq23}
z = \frac{2\omega_1}{\pi \theta_1'[0]} \frac{\sum_{m=1}^N \gamma_m \theta_1[m\beta/2] e^{(N-m)x}} {e^{Nx}} \qquad \genfrac{}{}{0pt}{}{\leftarrow \quad \mathcal{P}_{N-1}(e^x)}{} \:.
\end{equation}
The equation $z=0$ gives the solution \xpiu{} plus $N-1$ other solutions, 
which correspond to the roots of the polynomial $\mathcal{P}_{N-1}$. Now it is easy to organize the $N$ elliptic terms:
\begin{equation} \label{eq26}
G(x) = \sum_{m=1}^N \gamma_{N-m+1} \frac{ \thetauno{\dfrac{(N-m+1)\beta}{2}}{\tilde\tau} }{ \thetauno{\dfrac{m\beta}{2}}{\tilde\tau} } \frac{ \thetauno{ \dfrac{\pi z}{2\omega_1} + \dfrac{m\beta}{2} }{\tilde\tau} }{ \thetauno{\dfrac{\pi z}{2\omega_1}}{\tilde\tau} } \; e^{\textstyle mx}
\end{equation}
One may check the existence of a \hyph{(N+1)}order pole, and, thanks to \eqref{eq23},
the holomorphicity at the points $\{ x \text{ finite}, z=0 \}$.

\

Now we can expand $G(x)$ around \xpiu{} in powers of $e^x$, using for $z$ the expression
\eqref{eq23}, and imposing the behavior prescribed by $U(x)$. 
The calculation is quite long and does not give a closed expression, but it is algorithmic. 
For instance we have found the first parameter $\gamma_1$:
$$
\gamma_1 = - \left[ \sum_{i=1}^N e^{a_i} \right] \frac{\sin \frac{(N+1)\beta}{2}} {\sin \frac{N\beta}{2}} \: \frac{ \theta_1[\frac{N\beta}{2}] \, \theta_1'[0]} {\theta_1[\frac{N\beta}{2}] \, \theta_1'[\frac{\beta}{2}] + \theta_1[\frac{\beta}{2}] \theta_1'[\frac{N\beta}{2}]} \:,
$$
where $a_i$ are the roots of $V'(x)$. These are the same as the roots of $\sub{W'}{tree}(y)$, except for the change of variables \eqref{eq02}. This formula determines entirely the coefficient $\gamma_1$ as a function of known quantities, i.e. potential $\sub{W}{tree}$ and parameters $\beta$, $\lambda$, $M$. In this way we may find all the other parameters. Moreover, 
using \eqref{eq24}, we get the expectation value $\valmed{\tr \Phi}$ without integrating:
\begin{multline} \label{eq27}
\frac{\sin \frac{(N+1)\beta}{2}} {\sin \frac{N\beta}{2}} \: \frac{ \theta_1[\frac{N\beta}{2}] \, \theta_1'[0]} {\theta_1[\frac{N\beta}{2}] \, \theta_1'[\frac{\beta}{2}] + \theta_1[\frac{\beta}{2}] \theta_1'[\frac{N\beta}{2}]} \Bigl[ -\frac{\lambda}{M} \sum_{i=1}^N b_i + \frac{N}{2\sin \frac{\beta}{2}} \Bigr] = \\
\frac{1}{2\omega_1} \frac{ \theta_1'[0] \sin \frac{\beta}{2}}{\theta_1[\frac{\beta}{2}]} \Bigl[ -\frac{\lambda}{M} \langle \tr \Phi \rangle + \frac{N}{2 \sin \frac{\beta}{2}} \Bigr] \:,
\end{multline}
where $b_i$ are the roots of $W_{tree}(y)$. We chose to group factors in this manner because they tend to unit in 
both the semiclassical limit and $\beta \to 0$.

\subsection{The semiclassical limit}

The semiclassical limit corresponds to the weak coupling regime of the theory. 
We realize it by sending the physical coupling $g$ to zero, that is the complex (renormalized) coupling $\tau_R$ to $+i\infty$ or the $q$ parameter to zero. So the base torus covered by $\Sigma$ stretches to a cylinder, because the \hyph{B}period of $dz$ diverges and the cycle becomes non-compact. We expect the cuts in the $x$ space to shrink to singular points, that are really the critical points of $V(x)$. Meanwhile we expect the condensate fields to vanish, since they contain fermionic bilinear $\tr \lambda_{\alpha i} \lambda^\alpha_i$.

\

Since \eqref{eq18} is an absolutely convergent series, it is the most useful to take the limit $q \to 0$: we have to consider just the terms $k=0,-1$:
$$
e^{2i {\textstyle [ \frac{\pi z}{2\omega_1} - \frac{N\beta}{2} ]}} = \frac{H(x-i\frac{\beta}{2})} {H(x+i\frac{\beta}{2})} \:.
$$
To upper(lower) cuts on the cylinder, that is the points on the stretched torus which tend to $z=+i\infty(-i\infty)$, correspond to zeros of the denominator(numerator). The following map holds:
\begin{align*}
A_j : z\to +i\infty   \qquad &\Rightarrow \qquad \{ x=\xi_j + i\beta/2 \} \\
A_j^* : z\to -i\infty \qquad &\Rightarrow \qquad \{ x=\xi_j - i\beta/2 \} \:.
\end{align*}
This prove that the cuts $A_i$ and $A_i^*$ shrink to points that are the (shifted) roots of $H(x)$.

With the explicit expression \eqref{eq26} of $G(x)$ at hands, we can also prove that the chiral fields $S_i$ vanish. Expanding the $\theta$ functions for small $q$, and then considering the limit $z\to +i\infty$, $x \to \xi_j+i\beta/2$, we see that $G(x)$ converges to a constant in a neighborhood of a shrunk cut, 
therefore it is regular. So the integral $\oint_{A_i} G(x)dx$ which defines $S_i$ vanishes.

Finally, since the field $S$ vanishes, from \eqref{eq01} we get the support of $\rho(x)$:
$$
V'(x) = 0 \qquad x \in \supp \rho \:.
$$
So the singular points $\xi_j$ are really the critical points of $V(x)$:
the eigenvalues of $\Phi$ end up in the critical points. 
This expected behavior is confirmed by the semiclassical limit of \eqref{eq27}.

\acknowledgments

I am very grateful to Alberto Zaffaroni for having posed the problem and having given valuable suggestions, comments and remarks. I would like to thank also Jarah Evslin for useful discussion.

\end{document}